\documentclass[conference]{IEEEtran}
\IEEEoverridecommandlockouts
\usepackage{cite}
\usepackage{amsmath,amssymb,amsfonts}
\usepackage{algorithmic}
\usepackage{graphicx}
\usepackage{textcomp}
\usepackage{xcolor}
\usepackage{algorithm}
\usepackage{bm}
\usepackage{amsmath,amsfonts,amsthm}
\usepackage{mathtools}
\usepackage{makecell}
\usepackage{enumitem}
\usepackage{bm}
\usepackage{multirow}
\usepackage{url}
\usepackage{hyperref}
\usepackage[capitalize,noabbrev]{cleveref}
\usepackage{comment}
\usepackage{tikz}
\usepackage{soul}
\usepackage{fancyhdr}
\setul{1pt}{.4pt}
\def\BibTeX{{\rm B\kern-.05em{\sc i\kern-.025em b}\kern-.08em
    T\kern-.1667em\lower.7ex\hbox{E}\kern-.125emX}}

\makeatletter
\def\@IEEENORMtitlevspace{-1\baselineskip}
\def\@IEEEMINtitlevspace{-1\baselineskip}
\makeatother

\makeatletter
\def\@IEEENORMtitlevspace{0.25\baselineskip}
\def\@IEEEMINtitlevspace{0\baselineskip}
\makeatother

\newcommand{\algheader}[1]{%
  \begingroup\setlength{\topsep}{0pt}\setlength{\partopsep}{0pt}%
  \begin{flushleft}#1\end{flushleft}\endgroup}

\begin{document}

\definecolor{osuorange}{HTML}{E83600}

\newcommand{\Xin}[1]{{\textcolor{cyan}{\bf[[Xin: #1]]}}}
\newcommand{\Xuan}[1]{{\textcolor{blue}{\bf[[Xuan: #1]]}}}
\newcommand{\Wenbo}[1]{{\textcolor{osuorange}{\bf[[Wenbo: #1]]}}}

\title{Improving Progressive Compression with Adaptive Interpolation and Coefficient Decomposition
}
\makeatletter
\newcommand{\linebreakand}{%
  \end{@IEEEauthorhalign}
  \hfill\mbox{}\par
  \mbox{}\hfill\begin{@IEEEauthorhalign}
}
\makeatother

\author{
\IEEEauthorblockN{\parbox{2.1in}{\centering Wenbo Li}}
\IEEEauthorblockA{\parbox{2.1in}{\centering \textit{Oregon State University}\\ Corvallis, USA \\ liwenb@oregonstate.edu}}
\and
\IEEEauthorblockN{\parbox{2.1in}{\centering Xuan Wu}}
\IEEEauthorblockA{\parbox{2.1in}{\centering \textit{Oregon State University}\\ Corvallis, USA \\ wuxuan@oregonstate.edu}}
\and
\IEEEauthorblockN{\parbox{2.1in}{\centering Qian Gong}}
\IEEEauthorblockA{\parbox{2.1in}{\centering \textit{Oak Ridge National Laboratory}\\ Oak Ridge, USA \\ gongq@ornl.gov}}
\linebreakand
\IEEEauthorblockN{\parbox{2.1in}{\centering Pu Jiao}}
\IEEEauthorblockA{\parbox{2.1in}{\centering \textit{University of Kentucky}\\ Lexington, USA \\ pujiao@uky.edu}}
\and
\IEEEauthorblockN{\parbox{2.1in}{\centering Jieyang Chen}}
\IEEEauthorblockA{\parbox{2.1in}{\centering \textit{University of Oregon}\\ Eugene, USA \\ jieyang@uoregon.edu}}
\and
\IEEEauthorblockN{\parbox{2.1in}{\centering Qing Liu}}
\IEEEauthorblockA{\parbox{2.1in}{\centering \textit{New Jersey Institute of Technology}\\ Newark, USA \\ qliu@njit.edu}}
\linebreakand
\IEEEauthorblockN{\parbox{2.1in}{\centering Norbert Podhorszki}}
\IEEEauthorblockA{\parbox{2.1in}{\centering \textit{Oak Ridge National Laboratory}\\ Oak Ridge, USA \\ pnb@ornl.gov}}
\and
\IEEEauthorblockN{\parbox{2.1in}{\centering Scott Klasky}}
\IEEEauthorblockA{\parbox{2.1in}{\centering \textit{Oak Ridge National Laboratory}\\ Oak Ridge, USA \\ klasky@ornl.gov}}
\and
\IEEEauthorblockN{\parbox{2.1in}{\centering Xin Liang}}
\IEEEauthorblockA{\parbox{2.1in}{\centering \textit{Oregon State University}\\ Corvallis, USA \\ lianxin@oregonstate.edu}}
\thanks{This manuscript has been authored in part by UT-Battelle, LLC, under contract DE-AC05-00OR22725 with the US Department of Energy (DOE). The publisher, by accepting the article for publication, acknowledges that the U.S. Government retains a non-exclusive, paid up, irrevocable, world-wide license to publish or reproduce the published form of the manuscript, or allow others to do so, for U.S. Government purposes. The DOE will provide public access to these results in accordance with the DOE Public Access Plan
(http://energy.gov/downloads/doe-public-access-plan).}
}

\maketitle
\thispagestyle{fancy} 
\lhead{} 
\rhead{} 
\chead{} 
\lfoot{\footnotesize{ SC26, November 15-20, 2026, Chicago, Illinois, USA \newline 979-8-3195-4789-7/26/\$31.00 \copyright 2026 IEEE}}
\rfoot{} 
\cfoot{} 
\renewcommand{\headrulewidth}{0pt} 
\renewcommand{\footrulewidth}{0pt} 

\begin{abstract}
Exascale simulations generate data far faster than it can be stored or analyzed, making efficient data reduction essential. Error-controlled lossy compression offers high compression ratios under user-specified error bounds, but the target tolerance must be fixed at compression time. Progressive compression relaxes this restriction, yet existing methods still rely on fixed refactoring strategies and do not fully exploit correlations among decomposed coefficients, limiting the efficiency of progressive retrieval. 
In this work, we present an adaptive progressive compression framework that improves retrieval efficiency for two common targets, namely error-bound and peak Signal-to-Noise ratios. 
Our contributions are fourfold. (1) We propose to leverage two complementary interpolation schemes for adaptive progressive compression toward different targets, and we optimize them to achieve high efficiency. 
(2) We propose coefficient decomposition, a novel method that exploits the commonly overlooked spatial correlations among decorrelated data, which further improves the efficiency. 
(3) We develop the adaptive progressive compression workflow with automatic selection of the best-fit refactoring pipeline and tailored optimizations.
(4) We evaluate the proposed framework on five real-world scientific datasets against three state-of-the-art progressive compressors. 
Experimental results demonstrate that the proposed framework improves the compression ratio by up to $42.3\%$ under the same requested error tolerance and up to $92.5\%$ at the same PSNR, compared with the best-performing existing methods.
When transferring $512$ GB of scientific data to remote sites, the framework delivers up to $1.26\times$ speedup in the end-to-end data transfer performance.
Furthermore, our method achieves the highest visualization quality while retrieving the least amount of data from storage.
\end{abstract}


\section{Introduction}
Modern exascale simulations can generate data much faster than it can be stored, transferred, or analyzed. For example, a recent direct numerical simulation of isotropic turbulence on Frontier~\cite{frontier} produced about 0.5~PB per snapshot and would generate up to $\sim$38~TB/s if every time step were retained, far exceeding the 1--2 TB/s bandwidth of modern parallel file systems \cite{yeung2025jfm}. This growing gap makes efficient data reduction essential for computer and computational science.

Error-controlled lossy compression \cite{lakshminarasimhan2013isabela, sz17, lindstrom2006fast, lindstrom2014fixed, ainsworth2018multilevel} is a widely used approach to mitigate this bottleneck. Compressors such as \texttt{SZ} \cite{sz17, sz18, zhao2021optimizing, liang2022sz3}, \texttt{ZFP} \cite{lindstrom2014fixed}, and \texttt{MGARD} \cite{ainsworth2018multilevel, ainsworth2019multilevel, ainsworth2019qoi, liang2021mgard+} provide much higher compression ratios than lossless methods while guaranteeing that reconstruction error stays within a user-specified tolerance, and they have been integrated into scientific data management libraries such as \texttt{HDF5} \cite{hdf5} and \texttt{ADIOS-2} \cite{godoy2020adios}.
However, a key limitation of these single-error-bounded lossy compressors is that users must specify an error bound before storing the data.
Since the actual error tolerance required by various downstream analyses is often unknown a priori, and details lost during compression cannot be recovered, users tend to set conservative error bounds, thereby sacrificing compression ratio to preserve more information and reducing the overall effectiveness of lossy compression.

Progressive data compression~\cite{liang2021error, psz2025hpdc, bhatia2022amm, wallace1992jpeg, christopoulos2000jpeg2000, clyne2012progressive, li2026hpdc} tackles the above limitation by providing a more flexible reconstruction strategy. 
The idea originates from progressive image coding standards such as \texttt{JPEG} \cite{wallace1992jpeg} and \texttt{JPEG2000} \cite{christopoulos2000jpeg2000}, where data are refactored once into a near-lossless representation and then incrementally reconstructed until the desired accuracy is reached during retrieval.
These standards have long been used in modern browsers for progressive image rendering. However, they are not suitable for scientific applications because they cannot guarantee numerical error bounds.
To accommodate scientific use cases, \texttt{PMGARD}~\cite{liang2021error} combines the \texttt{MGARD} decomposition theory~\cite{ainsworth2019qoi} with bitplane encoding to offer guaranteed error control during progressive retrieval.
Nonetheless, its efficiency is limited by the multilinear decomposition it adopts and by its greedy retrieval strategy, both of which leave room for improvement.
Recently, \texttt{IPComp}~\cite{psz2025hpdc} achieves better efficiency by employing cubic spline interpolation and a dynamic programming-based retrieval strategy.
However, both \texttt{PMGARD} and \texttt{IPComp} are built around fixed interpolation pipelines, which limit their ability to adapt to different target error metrics and data characteristics. In addition, existing progressive compressors typically encode decomposed coefficients independently, overlooking their correlations and thereby leading to suboptimal efficiency. 

In this work, we design and develop an adaptive refactoring framework that improves retrieval efficiency for the two most commonly used types of target error tolerance. 
We further propose a novel method, namely coefficient decomposition, to improve efficiency by exploiting the spatial correlations within decomposed coefficients. 
We also develop an adaptive progressive compression workflow with online parameter tuning and tailored performance optimizations.
Our contributions are summarized as follows:
\begin{itemize}
    \item We leverage and optimize two complementary interpolation schemes for adaptive data decomposition in progressive compression, which offers high flexibility to achieve high efficiency towards diverse targets.   
    \item We propose a novel coefficient decomposition method that exploits spatial correlations within the decorrelated data after interpolation. This further improves retrieval efficiency with negligible metadata overhead. 
    \item We develop a fully adaptive progressive compression workflow that uses a sampling-based method to identify the best-fit refactoring pipeline on the fly and several tailored optimizations to mitigate overhead.
    \item We rigorously evaluate our method against state-of-the-art progressive compression methods through comprehensive experiments on five real-world datasets. Experimental results demonstrate that our method achieves competitive performance, the best retrieval efficiency, and good scalability, showing up to $42.3\%$ and $92.5\%$ improvement in compression ratio under the same error tolerance and PSNR, respectively, $1.26\times$ speedup in end-to-end transfer of $512$ GB data, and the highest visualization quality with the least retrieved data size.  
\end{itemize}

The rest of the paper is organized as follows. 
\cref{sec:related} discusses the related works. 
\cref{sec:overview} formulates the research problem and provides an overview of the proposed framework. 
\cref{sec:interpolation} introduces the adopted interpolation schemes.  
\cref{sec:coefficient} describes the proposed coefficient decomposition in detail. 
\cref{sec:implementation} summarizes the implementation and optimization. 
\cref{sec:evaluation} presents the experimental evaluation with state-of-the-art methods and real-world datasets. \cref{sec:conclusion} concludes the research with a vision for future work.
\section{Related Work}
\label{sec:related}

In this section, we recap related works on scientific lossy compressors and progressive compressors.

\subsection{Error-controlled Lossy Compressors}
Error-controlled lossy compressors target floating-point scientific data, where bounded numerical fidelity matters more than exact bitwise reconstruction. Unlike general-purpose lossless compressors such as \texttt{GZIP} \cite{gzip}, \texttt{ZSTD} \cite{zstd}, and \texttt{BLOSC} \cite{blosc}, they model spatial smoothness, multiscale structure, and value correlation in the data. 
Existing error-controlled compressors can be broadly grouped by how they decorrelate data before entropy coding. Prediction-based methods estimate each value from neighboring samples and then compress the residual. The \texttt{SZ} family \cite{sz17, sz18, zhao2021optimizing, liang2022sz3} is the best-known representative in this category. Early versions of \texttt{SZ} center on a Lorenzo predictor \cite{ibarria2003out}, while later versions incorporate richer models such as regression and interpolation-based predictors \cite{sz18, zhao2021optimizing}. After prediction, the residuals are quantized with a linear-scaling quantizer \cite{sz17} and compressed through entropy encoding, such as Huffman coding \cite{huffman1952method}, together with lossless coders like \texttt{ZSTD} \cite{zstd}.
Transform-based compressors instead map the data into another basis in which a small set of coefficients captures most of the data content. \texttt{ZFP} \cite{lindstrom2014fixed} is a representative design that partitions data into small blocks, converts values to fixed-point form under a common exponent, applies a near-orthogonal transform, and then uses embedded coding to emit only as many transformed bits as needed to satisfy requested accuracy.
\texttt{MGARD} \cite{ainsworth2018multilevel, ainsworth2019multilevel, ainsworth2019qoi} provides another important line of work. It combines multilevel decomposition with finite-element analysis and derives error control from a mathematically grounded hierarchy.

Despite their methodological differences, the prevailing usage model of these compressors remains the same, where compression is performed against a single user-specified tolerance, and the compressed representation is tailored to satisfy that tolerance alone.
In practice, however, the same dataset often serves multiple downstream analyses with varying precision requirements. Since lost information cannot be recovered after decompression, users are compelled to select a conservatively tight error bound that satisfies the most demanding analysis, leading to unnecessarily low compression ratios and diminished benefits in data storage and transfer.

\subsection{Data Refactoring and Progressive Retrieval}
Progressive compression studies not only how data are compacted, but also how compressed information is organized so that reconstruction can proceed progressively. In this setting, a compressed representation is usually divided into ordered components such as bitplanes and hierarchical levels, so that it can be used to incrementally reconstruct the data to any desired accuracy during retrieval.
This write-once, retrieve-progressively strategy is particularly well-suited to scientific data management, where a dataset is generated once and subsequently consumed by numerous analyses with different fidelity requirements.

The roots of this idea can be traced to progressive image coding methods such as \texttt{JPEG} and \texttt{JPEG2000} \cite{wallace1992jpeg, christopoulos2000jpeg2000}, which organize encoded information so that visual quality improves gradually as more bits arrive. Their influence on scientific compression lies mainly in progressive transmission and embedded coding concepts.

\texttt{PMGARD} \cite{liang2021error} is among the first compressors to bring this progressive retrieval paradigm into scientific error-bounded compression. Built on the \texttt{MGARD} theory \cite{ainsworth2019multilevel, ainsworth2019qoi}, it organizes multilevel coefficients into bitplane streams that can be decoded in stages while preserving strict error guarantees for both $L^2$ and $L^{\infty}$ targets. However, \texttt{PMGARD} relies on linear interpolation during decomposition, which limits its ability to capture data correlations compared to higher-order interpolation schemes. Its greedy retrieval strategy also does not minimize retrieval volume for a given tolerance. Additionally, although \texttt{PMGARD} supports different target error metrics through per-bit and negabinary encodings, these changes are largely confined to the encoding phase, while the underlying decorrelation method remains the same.

Magri et al. \cite{magri2023general} explore a simpler construction that builds a progressive representation by repeatedly invoking an existing error-controlled lossy compressor with a sequence of decreasing tolerances. At a high level, the original data is first compressed at a coarse tolerance, then finer residuals are compressed successively to form additional refinement levels. While straightforward to implement, its drawback is that retrieval becomes accumulation-heavy, since reconstructing a fine level requires decoding all preceding levels and summing their contributions, causing the retrieval cost to grow with the number of refinement stages.

\texttt{IPComp} \cite{psz2025hpdc} advances the state of the art by combining the interpolation-based decomposition from \texttt{SZ3} \cite{zhao2021optimizing} with a dynamic-programming retrieval strategy. Compared with \texttt{PMGARD}, its cubic spline interpolation captures stronger local correlation, and its retrieval algorithm is better aligned with minimizing retrieved data volume. Nevertheless, notable limitations remain. 
First, it quantizes the data with a preset error bound before decorrelation, thereby reducing the achievable precision and degrading decorrelation accuracy.  
Second, it leverages a fixed pipeline that cannot adapt to different error metrics. 
Last but not least, it overlooks the correlation in the decorrelated coefficients as existing work, which limits the overall efficiency. 


These shortcomings motivate the present work, in which we design an adaptive refactoring framework that addresses the above limitations through adaptive interpolation and encoding, a novel coefficient decomposition method that exploits spatial correlations within decomposed data, and a systematic design that delivers an adaptive progressive compression pipeline with tailored performance optimizations. 
\section{Overview}
\label{sec:overview}

In this section, we formulate the research problem and provide an overview of the proposed framework.

\subsection{Problem Formulation}
We define the error-controlled progressive method, and then provide a formal formulation of the target-driven data refactoring and retrieval. 
An error-controlled progressive method refactors the original data $\bm{X}=\{x_1, \dots, x_n\}$ into $k$ progressive segments $\{s_1, \dots, s_{k}\}$. 
Given a user-requested error bound $\tau$ during retrieval, the method is supposed to automatically determine the number of fragments required (denoted as $n_\tau$), and use them to reconstruct a decompressed data $\bm{X'}=\{x_1',\dots,x_n'\}$ which satisfies $\max_i |x_i - x_i'| \leq \tau$.
The total retrieved size under $\tau$ is defined as $S_\tau=\sum_{i=1}^{n_\tau} \text{size}(s_i)$.

We identify two related but different targets for progressive retrieval based on the diverse needs of scientific applications: 
(1) error bound and (2) peak Signal-to-Noise ratio (PSNR). 
An error bound limits the maximal error in the reconstructed data, which is needed in applications that require guaranteed absolute error control, as mentioned in many error-controlled lossy compressors~\cite{sz17, lindstrom2014fixed, ainsworth2018multilevel}. 
PSNR is defined as $20\log_{10}{(\max_i{x_i} - \min_i{x_i})/RMSE}$, where $RMSE=\sqrt{\sum_{i=1}^{n}(x_i - x_i')^2/n}$ denotes the root of mean squared errors. 
It measures the average error over the entire domain, and this is preferred in applications that care about the global error distribution, which is common in fusion energy science and climate studies~\cite{gong2021maintaining, baker2016evaluating}. 
While both of them reflect the quality of the reconstructed data, there is no universal solution that optimizes both. 
Accordingly, we formulate our two targets separately as
\begin{equation*}
    \min S_{\tau} \quad \text{s.t.,} \quad \|\bm{X} - \bm{X'}\|_{\infty} \leq \tau, 
\end{equation*}
for error-bounded retrieval, and
\begin{equation*}
    \max \; \mathrm{PSNR}(\bm{X}, \bm{X'}) \quad \text{s.t.,} \quad S = S_{\tau}, \; \|\bm{X} - \bm{X'}\|_{\infty} \leq \tau,    
\end{equation*}
for PSNR-oriented retrieval.
In the rest of the paper, we refer to our designs toward the two targets as the error-bound mode and the PSNR mode, respectively.


\subsection{Overview}
We present an overview of the proposed framework in \cref{fig:overview}, with blue boxes representing proposed components and cyan boxes indicating optimized components, and gray boxes denoting components adjusted accordingly in scientific data refactoring and progressive retrieval pipelines.
Similar to existing methods, our framework has a data refactoring stage, which writes data upon generation, and a retrieval stage, which retrieves data with guaranteed error control for post hoc data analytics.
During data refactoring, we leverage and optimize two complementary interpolation schemes and encoding methods to enable an adaptive, flexible refactoring pipeline for diverse targets.
We also propose coefficient decomposition, which exploits the commonly overlooked spatial correlation within the decorrelated coefficients to improve retrieval efficiency.
During retrieval, we adjust the retrieval estimation, bitplane decoding, and interpolation components to incorporate the adaptive interpolation schemes and the proposed coefficient decomposition. 

\begin{figure}[t]
    \centering
    \includegraphics[width=\linewidth]{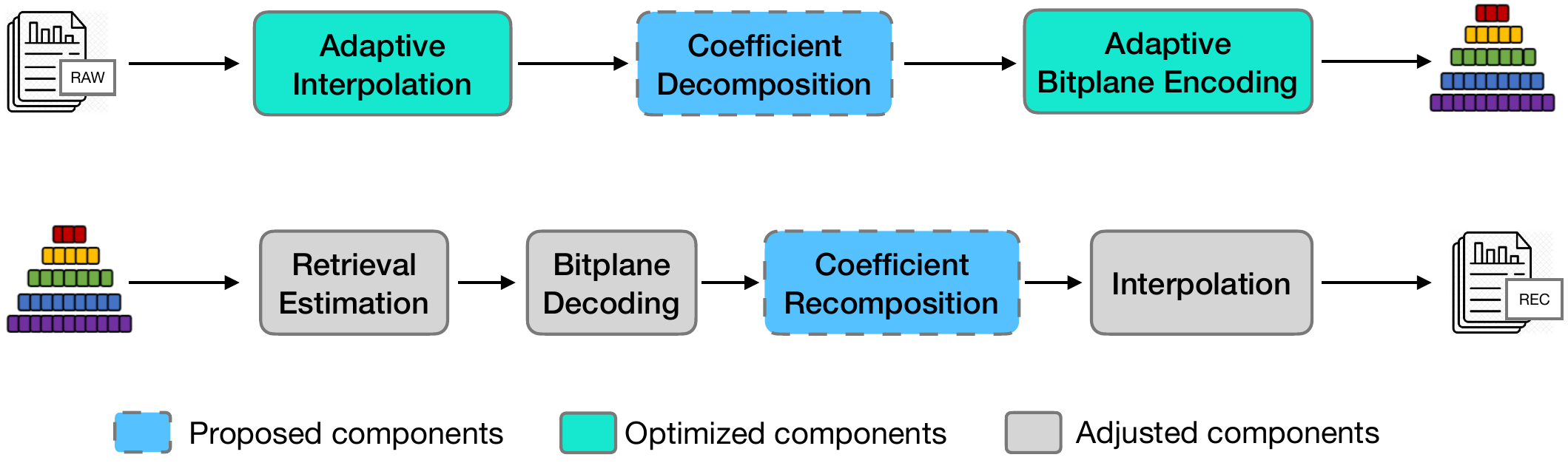}
    \caption{Overview of the proposed framework. 
    }
    \label{fig:overview}
\end{figure}

\section{Adaptive Interpolation}\label{sec:interpolation}

We leverage two multilevel interpolation schemes in our framework to adapt to diverse targets, each with distinct characteristics. 
We refer to them as per-level interpolation and per-region interpolation, respectively, based on how they interpolate data within a level. 
In this section, we introduce the two schemes and analyze their efficiency with respect to the two targets (error bound and PSNR). 

Both interpolation schemes decompose the data into levels with different strides in a bottom-up fashion. 
Levels are used to represent different subsets of data points in an embedded hierarchy, as detailed below.  
The decomposition starts with a stride of 1 at the finest level (level 0), where all data points are present. 
For any specific level $l$, it contains only data points in level $l-1$ with even indices along each dimension, i.e., it reduces the resolution by half and doubles the stride in each dimension. 
The decomposition produces $L$ levels based on user inputs, and terminates at the level $L-1$, which is typically a coarse representation that accounts for a very small fraction of the data. 

While the two schemes share the same multilevel decomposition, they feature distinct intra-level interpolation methods for data decorrelation.  
The procedures for linear interpolation with 3D datasets are illustrated in \cref{fig:interp}, and they generally extend to any arbitrary dimensionality and higher-order interpolation methods. 
We describe the two schemes below and highlight our optimizations relative to existing work. 

\begin{figure}[htbp]
    \centering
    \includegraphics[width=\linewidth]{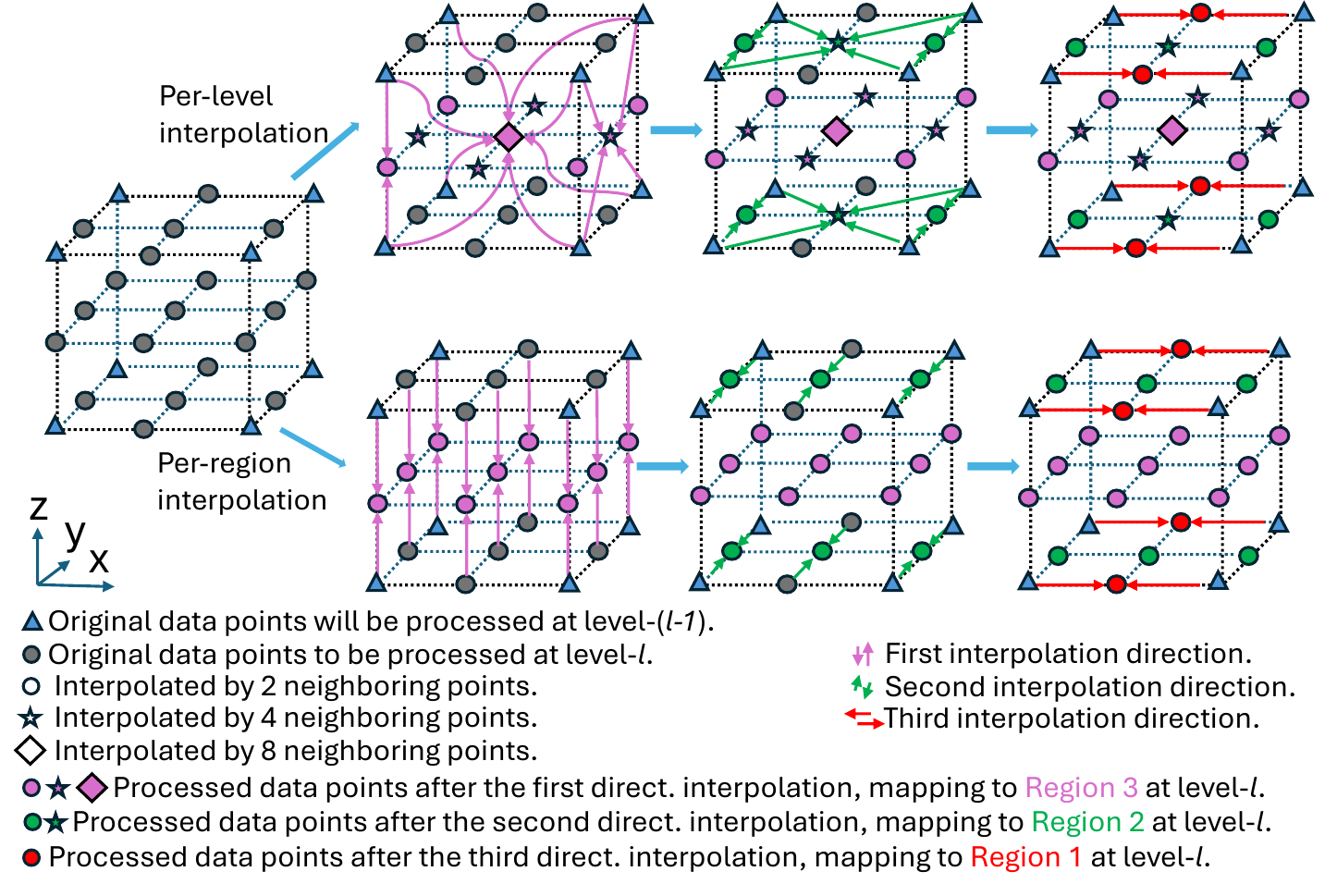}
    \caption{Illustration of the decomposition pipelines of per-level linear interpolation and per-region linear interpolation.}
    \label{fig:interp}
\end{figure}

\textbf{Per-level interpolation:} The per-level interpolation scheme originates from \texttt{PMGARD}~\cite{liang2021error}, where the intra-level interpolation uses only data points from a higher level. 
As shown in the top row of \cref{fig:interp}, it uses the eight data points in the upper level to interpolate all remaining data points in the current level with multilinear interpolation. 
As such, the data points at the centers of the edges, faces, and cubes are interpolated using the averages of 2, 4, and 8 corresponding upper-level data points, respectively. 
This scheme has limited error propagation because errors at the current level are completely independent. 
However, it has relatively low interpolation accuracy because data points at the center of faces and cubes are interpolated using faraway upper-level data points. 
Compared to \texttt{PMGARD}~\cite{liang2021error}, our per-level interpolation features tricubic interpolation for higher efficiency, and it directly interpolates the data with strides to avoid the expensive reordering steps for higher throughput. 

\textbf{Per-region interpolation:} The per-region interpolation scheme is proposed in SZ3~\cite{zhao2021optimizing} and extended for data decomposition in progressive compression in \texttt{IPComp}~\cite{psz2025hpdc}. 
Instead of using fixed data points for interpolation, it adopts a fixed mechanism: each data point in the current level is interpolated using the same formula (e.g., average of two points in the linear example), but with different data. 
This is typically done by performing separate interpolations along each dimension, as noted in the bottom row of \cref{fig:interp}. 
We name all data points interpolated along the same dimension a ``region", so this scheme is called per-region interpolation. 
Since this scheme uses only adjacent data points for interpolation, it tends to achieve higher accuracy. 
However, it suffers from error propagation, as errors can propagate across regions at the same level during reconstruction (e.g., from region one to region two when interpolating the centering data point on the top face).
Compared to \texttt{IPComp}~\cite{psz2025hpdc}, our per-region interpolation features a bottom-up design and omits the quantization stage, thereby providing more accurate decorrelation using the original data values rather than the quantized ones. 
We also use a tuner to identify the best-fit interpolation orders across different dimensions, as prior studies~\cite{zhao2021optimizing, liu2024high} show that these orders can affect both interpolation accuracy and throughput.

Based on the above analysis, we found that per-level interpolation is preferred for the error-bound mode and per-region interpolation is favorable for the PSNR mode for most cases. 
As such, we advocate an adaptive interpolation scheme that automatically adjusts based on the target. In practice, we use a tuner to determine which interpolation scheme to use on the fly, as will be detailed in \cref{sec:implementation}.

\section{Coefficient Decomposition}\label{sec:coefficient}
Coefficients are the residues between original data values and their interpolated counterparts using the schemes in \cref{sec:interpolation}. 
They are generally encoded using bitplane encoding and lossless compression, and their distribution is a key factor in determining the efficiency of progressive compression: a large percentage of near-zero coefficients indicates a high percentage of zeros in the most significant bitplanes and thus higher compressibility. 
Inspired by prior studies~\cite{jiao2025qp} that investigated the correlation of quantization indices in error-controlled lossy compressors, we propose an efficient coefficient decomposition method that leverages coefficient correlations for better progressive compression. 

\begin{figure}[t]
    \centering
    \includegraphics[width=\linewidth]{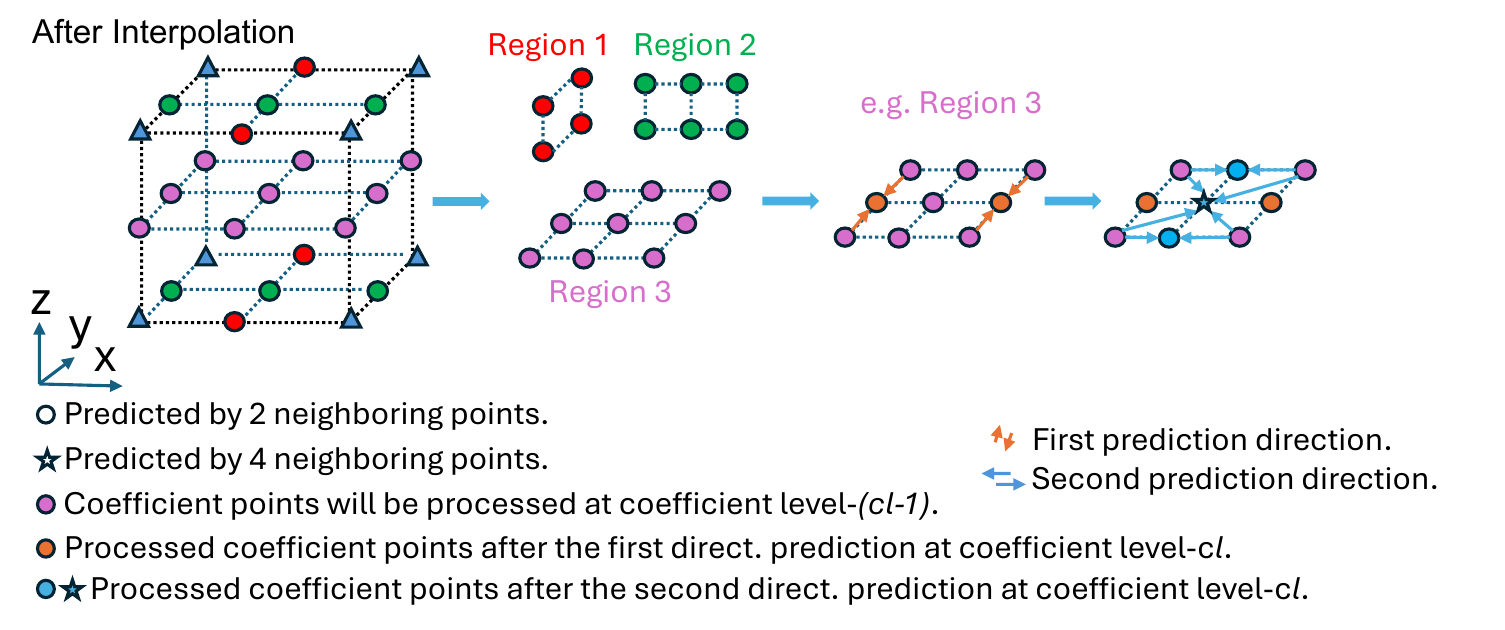}
    \caption{Illustration of the 2D coefficient decomposition with linear interpolation on 3D data.}
    \label{fig:coefficient_interp}
\end{figure}

\begin{figure}[t]
    \centering
    \includegraphics[width=\linewidth]{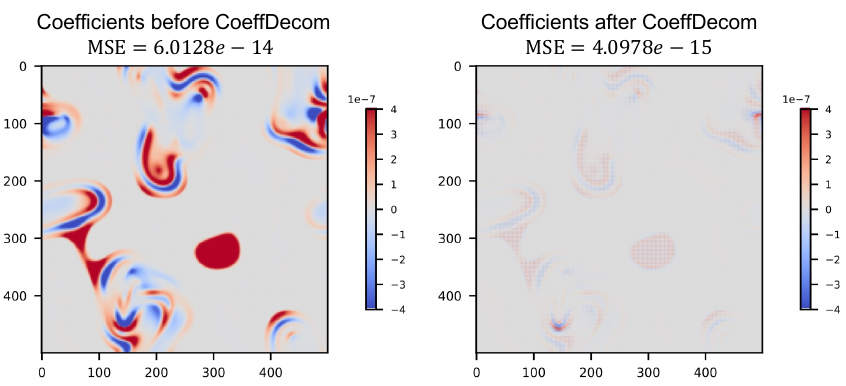}
    \caption{Slice $[:, :, 125]$ of the $500\times500\times250$ coefficients in Region 3 at the finest level from the \textit{CH\textsubscript{4}} field of the S3D dataset, decomposed using cubic adaptive interpolation. Left: original coefficients. Right: coefficients after 2D linear CoeffDecom.}
    \label{fig:coefficient_per-level}
\end{figure}

\subsection{Motivation}

Following the procedure in~\cite{jiao2025qp}, we group the coefficients in different regions into separate groups, as shown in \cref{fig:coefficient_interp}. We then extract a 2D slice in Region 3 and visualize it in \cref{fig:coefficient_per-level} using the \textit{CH\textsubscript{4}} fields of the S3D dataset as an example (see \cref{tab:data} for dataset information). 
According to this figure, coefficients in this region exhibit high correlation, and similar phenomena are observed in Region 1 and Region 2. 
In addition, this phenomenon occurs in both per-level interpolation and per-region interpolation, because the residuals tend to be smooth as they are the differences between two smooth sets of values: the original scientific data and the interpolated data. 
These observations motivate us to further decorrelate the coefficients for better efficiency. 

\subsection{Methodology}
Prior studies~\cite{jiao2025qp} use a Lorenzo predictor to decorrelate the quantization indices in error-controlled lossy compressors, but this approach cannot generalize to progressive compression. 
This is because the Lorenzo predictor exhibits unbounded error propagation, which is not a problem for quantization indices with a fixed value (i.e., errors are all 0) but matters for coefficients whose values vary with the number of retrieved bitplanes. 
Instead, we propose to further decompose the coefficients using per-level interpolation, as detailed below. 


The latter half of \cref{fig:coefficient_interp} depicts our coefficient decomposition method on Region 3, with coefficients computed from per-region interpolation, and the same procedure applies for other regions and per-level interpolation. 
In particular, we treat the 3D coefficients as a stack of multiple 2D slices, and perform 2D per-level multilinear interpolation on each slice. 
This is done on each region to ensure all the correlations are properly handled. 
We use 2D interpolation instead of 3D interpolation because such correlations are mainly seen in the hyperplane orthogonal to the interpolation direction, which also aligns with the observations in~\cite{jiao2025qp}. 
We use the per-level scheme and multilinear interpolation because they have minimal error propagation, and we fix the total number of levels to three for the same reason. 
The decomposed coefficients for the same data are visualized in \cref{fig:coefficient_per-level}. It is clearly observed that they have more near-zero data points with much smaller mean squared errors. 

The coefficient decomposition method can be applied at any level, but we observe a noticeable improvement mainly at the finest level. 
This is because coefficients at the higher levels (1) only take up a small percentage of data (e.g., less than 1/8 for 3D data); and (2) have weaker correlation as they are spatially farther away from each other. 
As such, we only perform coefficient decomposition at the finest level throughout the paper. 

\subsection{Coefficient storage and retrieval}\label{sec:storage}
While coefficient decomposition reduces entropy in the coefficients, it introduces new challenges for retrieval due to interleaved regions and added levels. 
In what follows, we detail our adjustments to ensure error-controlled retrieval when coefficient decomposition is coupled with the two interpolation schemes in \cref{sec:interpolation}. 

\paragraph{Per-level interpolation} In the original design, one level of coefficients in per-level interpolation is encoded into bitplanes as a whole. 
The maximum value of the coefficients is stored as metadata and used to establish 1-to-1 mappings from bitplane to absolute errors, which provide essential information for error-controlled retrieval algorithms (greedy-based in \texttt{PMGARD}~\cite{liang2021error} and dynamic programming in \texttt{IPComp}~\cite{psz2025hpdc}). 
With coefficient decomposition, such mappings cannot be established directly because the actual absolute error at the level is the maximum error across all regions.

\begin{figure}[htbp]
    \centering
    \includegraphics[width=\linewidth]{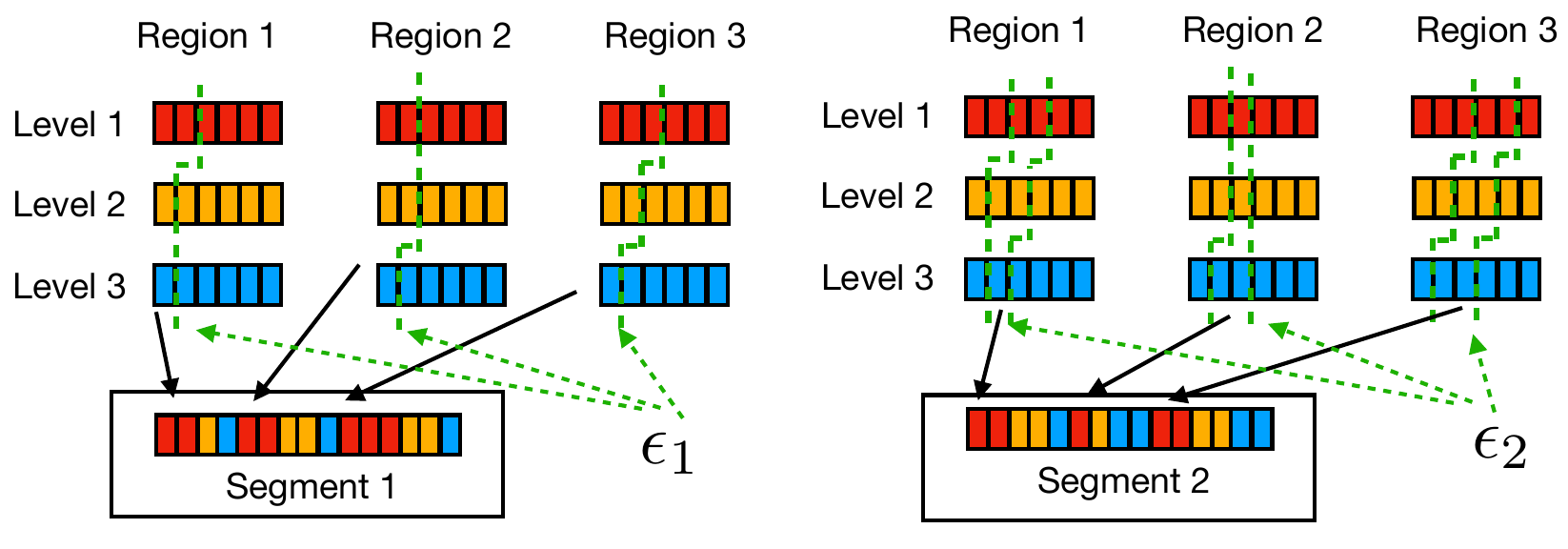}
    \caption{Establishing segment-error bound mappings across three regions using two identified error bounds. A small rectangle represents a bitplane in the corresponding level, and our algorithm merges bitplanes across regions to form segments enforcing a set of target error bounds. }
    \label{fig:eb_merge}
\end{figure}
To address this issue, we couple the greedy-based retrieval algorithm in~\cite{liang2021error} and a max ordering mechanism to establish such mappings. 
In particular, we first leverage the greedy-based algorithm to determine the loading order of bitplanes and their corresponding error bounds within each region, and then identify a set of similar error bounds across all regions.
After that, we apply \textit{max ordering} to arrange the bitplanes across all the regions at the level.
Since the regions contribute equally to the global error bound through the $\max$ operation, we merge their orderings by sorting all steps across the three regions in descending order of their cumulative error.
When multiple steps from different regions have the same cumulative error, they are merged into a single step.
This results in a unified loading order for bitplanes across all regions at the level, along with the corresponding $\max$ error after each step. 
\cref{fig:eb_merge} demonstrates how this algorithm works with two sample error bounds $\epsilon_1$ and $\epsilon_2$, and it generally extends to any number of error bounds.  
In particular, our algorithm identifies $\epsilon_1$ and $\epsilon_2$ as two target error bounds across regions, and then it merges bitplanes across all regions to ensure the merged segment enforces these error bounds. 
To this end, a list of 1-to-1 mappings from segments to error bounds is established and all the error bounds are stored as metadata. 
During retrieval, such information is used to enable guaranteed error control. 
This increases the metadata overhead, since one level now records a list of error bounds rather than a single maximum value. Nonetheless, this overhead remains negligible relative to the data size. 

\paragraph{Per-region interpolation} Since errors propagate across regions and levels in the same way, prior works~\cite{psz2025hpdc} apply the retrieval algorithms at the granularity of regions. This yields $3(L-1)+1$ regions to process, given $L$ levels in total.   
Since our regions are the same as those in per-region interpolation, we directly split any region with coefficient decomposition to three new regions, each of which corresponds to a decomposed coefficient level. 
After that, we can directly apply existing algorithms, such as the one in~\cite{psz2025hpdc}, with only minimal modification to the error estimation mechanisms (i.e., changing the error estimation from cubic interpolation to linear interpolation in the decomposed coefficient regions).

\section{Implementation and Optimization}\label{sec:implementation}
In this section, we introduce our detailed implementation of the adaptive progressive compression pipeline. 
We first present our target-driven tuning for automatic component selection, followed by a detailed algorithm along with tailored performance optimizations. 

\subsection{Target-driven tuning}
We leverage a target-driven tuning method to determine the configuration for our data refactoring pipeline online.
In particular, we uniformly sample $1\%$ of the original data and evaluate the retrieval efficiency of each option across 9 commonly used tolerances which are $10^{-3}$, $5\times10^{-4}$, $10^{-4}$, $\dots$, $10^{-7}$. 
To this end, we employ a voting scheme to identify the best-fit configuration, i.e., the one that receives the most votes in the trials. 
We also conduct an offline study to justify the $1\%$ sampling ratio, comparing the configuration voted from the sample against the one voted from the full data across all fields of all datasets for a single-machine experiment.
Experimental results demonstrate that sampled tuning reproduces the full-data decision for $85.0\%$ of the interpolation-scheme selections and $88.1\%$ of the coefficient-decomposition selections, while reducing the tuning overhead from $87.2\%$ of the total refactoring time down to $11.5\%$.
In addition, the residual disagreements are largely benign.
Evaluating the selected configurations over a wider range of 16 tolerances, from $10^{-1}$ to $10^{-9}$, i.e., beyond the range on which the vote is taken, the resulting bitrate differs by only $1.38\%$ on average over the cases where the two tuners diverge, and by $0.37\%$ when averaged over all cases.
At the loose tolerances outside the voting range, neither tuner is optimized, and the sampled tuner occasionally yields a slightly \emph{lower} bitrate.
We therefore adopt $1\%$ sampling for online tuning.

\begin{table}[htbp]
{
\footnotesize
\centering
\caption{Tuning and \ul{tuned} configuration for 3D data}
\label{tab:tuning}
\footnotesize
\resizebox{\columnwidth}{!}{%
\begin{tabular}{|l|c|c|c|c|}
\hline
\thead{} & \thead{Error-bound mode} & \thead{PSNR mode} \\ 
\hline
Interpolation schemes & Per-level / Per-region (6) & Per-region (6) \\
\hline
Coefficient decomposition & No / 3 directions & No / 3 directions \\
\hline
Encoding & \ul{Generic bitplane} & \ul{Negabinary with XOR} \\
\hline
\end{tabular}
}
}
\end{table}

Table~\ref{tab:tuning} summarizes the tuning configurations in our framework. 
In particular, we will evaluate different interpolation schemes, with per-region interpolation comprising 6 variants that represent different permutations of the interpolation order.  
For coefficient decomposition, we evaluate four options, including three for performing decomposition along 3 different directions and one for skipping. 
The underlined configurations in the table are called tuned, as they consistently exhibit higher efficiency than their alternatives in offline studies. 
In particular, we found that (1) the generic bitplane encoding is always better than negabinary encoding in the error-bound mode but worse in PSNR mode, and (2) including XOR with negabinary encoding always leads to better efficiency. 
These findings align with the design choices in prior work~\cite {liang2021error, psz2025hpdc}. In addition, we found that per-region interpolation is consistently better than per-level interpolation in PSNR mode, and therefore omit the latter for online tuning.

\subsection{Algorithm}
We introduce our refactoring and retrieval algorithm with per-level interpolation for demonstration purposes, and a similar procedure applies to per-region interpolation. 
Algorithm~\ref{alg:refactoring} presents our refactoring algorithm. 
We initialize the current stride to 1 in the beginning and perform interpolation at the finest level (lines 1-2). 
Afterward, we interleave the data into different regions to perform coefficient decomposition (lines 3-16).
In particular, we first iterate and process coefficients in each region, and we treat them as a stack of multiple slices (lines 5-7).
Coefficients in each slice are decomposed into three levels and collected separately (lines 9-12). 
When all the slices are processed in the region, bitplane encoding is performed on all the levels to produce binary streams (line 14).
After all the regions are processed, the max ordering strategy (see \cref{sec:storage}) is used to merge bitplanes to segments with metadata $M_0$ denoting the segment to error bound mappings (line 16).
This completes the coefficient decomposition, and the remaining procedure for refactoring later levels is the same as the one in \texttt{PMGARD}~\cite{liang2021error}, except that cubic interpolation is used to achieve better efficiency. 

\begin{algorithm}[t]
\caption{Refactoring with Coefficient Decomposition} \label{alg:refactoring} \footnotesize
\renewcommand{\algorithmiccomment}[1]{/*#1*/}
\algheader{%
\textbf{Input}: input data $X$, number of levels $L$\\
\textbf{Output}: bitplanes $\{bp_i\}$ and metadata $\{M_i\}$}
\begin{algorithmic} [1]
\STATE $stride \gets 1$ \COMMENT{Initialize stride}
\STATE $\Pi \gets $ \texttt{CubicInterpolate}($X, stride$) \COMMENT{Perform interpolation}
\STATE $regions \gets$ \texttt{Interleave}($X - \Pi, stride$) \COMMENT{Interleave regions}
\STATE $bp \gets \emptyset$ \COMMENT{Initialize the set to store region bitplanes}
\FOR{$region \in regions$}
    \STATE \COMMENT{Iterate and process each region}
    \FOR{$S \in region$} 
        \STATE \COMMENT{Iterate and decompose each slice}
        \STATE $\Pi_1 \gets $\texttt{LinearInterpolate}($S, 1$)  \COMMENT{Interpolate with stride 1}
        \STATE $\Pi_2 \gets $\texttt{LinearInterpolate}($S, 2$)  \COMMENT{Interpolate with stride 2}
        \STATE $S_1, S_2, S_3 \gets $ \texttt{ExtractLevel}($S - \Pi_1 - \Pi_2$) \COMMENT{Extract decomposed coefficients based on their levels}
        \STATE $R_1 \gets R_1 \cup S_1$, $R_2 \gets R_2 \cup S_2$, $R_3 \gets R_3 \cup S_3$ \COMMENT{Collect decomposed coefficients based on their levels}
    \ENDFOR
    \STATE $bp \gets bp \cup $ \texttt{BitplaneEncoding}(R1, R2, R3) \COMMENT{Encode decomposed coefficients on all levels}
\ENDFOR
\STATE $bp_0, M_0 \gets $\texttt{MaxOrdering}($bp$) \COMMENT{Establish segment-error bound mapping, see \cref{sec:storage}}
\FOR{$l = 1 \to L-1$}
    \STATE $stride \gets 2 * stride$
    \STATE $\Pi \gets $ \texttt{CubicInterpolate}($X, stride$)
    \STATE $bp_l, M_l \gets$ \texttt{BitplaneEncoding}($X - \Pi, stride$)
\ENDFOR

\RETURN $\{bp_0, bp_1, \cdots, bp_{L-1}\} \cup \{M_0, M_1, \cdots, M_{L-1}\}$
\end{algorithmic}
\end{algorithm}

We then present the corresponding retrieval algorithm in Algorithm~\ref{alg:retrieval}. 
We start with the largest stride because reconstruction has to start from the coarsest level (line 1). 
After initializing reconstructed data (line 2), we determine which bitplanes to retrieve using an interpretation algorithm and the metadata.
For per-region interpolation, we directly use the dynamic programming-based method in~\cite{psz2025hpdc}, as it delivers better efficiency with negligible performance overhead. 
For per-level interpolation, we cannot use the method directly because the merged segments have a different format in metadata. 
As such, we use the best-first search algorithm with pruning and incremental invocation to find the appropriate bitplanes with similar efficiency, but it incurs slightly higher overhead. 
After the bitplanes are identified and fetched, we reconstruct the data from level $L-1$ to $1$ as in the existing approach (lines 4-9). 
At the finest level, we reconstruct the slices within each region one by one and merge them to form the regions (lines 10-20). 
In the end, we restore the regions to their corresponding locations and add them back to the data after the final interpolation (lines 22-24). 

\begin{algorithm}[t]
\caption{Retrieval with Coefficient Decomposition} \label{alg:retrieval} \footnotesize
\renewcommand{\algorithmiccomment}[1]{/*#1*/}
\begin{flushleft}
\textbf{Input}: requested error bound $\tau$, bitplanes $\{bp_i\}$ and metadata $\{M_i\}$\\
\textbf{Output}: reconstructed data $X'$
\end{flushleft}
\begin{algorithmic} [1]
\STATE $stride \gets 2^{L}$ \COMMENT{Initialize stride}
\STATE $X' \gets \bm{0}$ \COMMENT{Initialize reconstructed data}
\STATE $\{bp_i\} \gets$ \texttt{RetrievalInterpreter}($\tau$, $\{M_i\}$) \COMMENT{Determine how many bitplanes to retrieve}
\FOR{$l = L-1 \to 1$}
    \STATE $stride \gets stride / 2$
    \STATE $C \gets$ \texttt{BitplaneDecoding}($\{bp_i\}$) \COMMENT{Decode coefficients}
    \STATE $\Pi \gets $ \texttt{CubicInterpolate}($X, stride$) \COMMENT{Perform Interpolation}
    \STATE $X' \gets \Pi + $ \texttt{ExpandByStride}($C, stride$) \COMMENT{Add coefficients back}
\ENDFOR
\FOR{$region \in regions$}
    \STATE $region \gets \emptyset$ \COMMENT{Initialize region}
    \STATE $R_1, R_2, R_3 \gets$ \texttt{BitplaneDecoding}($bp_0$) \COMMENT{Decode bitplanes for all levels in region}
    \FOR{$S \in region$} 
        \STATE \COMMENT{Iterate and recompose each slice}
        \STATE $S_1, S_2, S_3 \gets $\texttt{GetSlice}($R_1, R_2, R_3$) \COMMENT{Get slice from decoded region}
        \STATE $\Pi_1 \gets $\texttt{LinearInterpolate}($S_3, 1$)  \COMMENT{Interpolate level 2}
        \STATE $\Pi_2 \gets $\texttt{LinearInterpolate}($S_2, 2$)  \COMMENT{Interpolate level 1}
        \STATE $S \gets $ \texttt{ExpandByStride}($S_1, 4$) + \texttt{ExpandByStride}($S_2, 2$) + $S_3$ + $\Pi_1$ + $\Pi_2$ \COMMENT{Reconstruct the slice}
        \STATE $region \gets region \cup S$ \COMMENT{Merge slice to region}
    \ENDFOR
    
\ENDFOR
\STATE $C \gets$ \texttt{Reposition}($regions$) \COMMENT{Put recomposed coefficient back to correct locations}
\STATE $\Pi \gets $ \texttt{CubicInterpolate}($X', stride$) \COMMENT{Perform interpolation}
\STATE $X' \gets \Pi + C$  \COMMENT{Add recomposed coefficients back}
\RETURN $X'$
\end{algorithmic}
\end{algorithm}

We fix the decomposition depth to three levels based on offline studies. 
Fewer levels leave residual spatial structure in the coefficients under-exploited, whereas deeper hierarchies propagate error across more levels. 
To quantify this trade-off, we compare the three depths on the SCALE dataset at 16 error bounds spanning $10^{-1}$ to $10^{-9}$, averaging the bit-rate over 11 fields at each. 
At every error bound, we take the lowest of the three bit-rates as the reference and measure the excess of each depth over it. 
For instance, at $\epsilon=10^{-5}$ in the error-bound mode, three-level yields the lowest bit-rate, while two-level and four-level exceed it by $1.1\%$ and $0.9\%$, respectively.

Table~\ref{tab:depth} reports, for each depth, the number of error bounds at which it is the cheapest (\emph{\#best}) together with its average and worst
excess across the 16 error bounds. 
No single depth wins everywhere, but the penalty for a wrong choice is strongly asymmetric. 
Three-level is the cheapest at 11 of the 16 error bounds in the error-bound mode and at 13 in the PSNR mode, and whenever it loses it does so by at most $0.83\%$. 
Four-level gives up as much as $3.71\%$ where it is not preferred, and two-level never wins at all while costing up to $5.00\%$. 
The error bounds favoring four-level also differ between the two modes, so no deeper configuration is uniformly preferable.
According to these observations, we adopt three as the decomposition depth throughout the paper. 

\begin{table}[htbp]
{
\footnotesize
\centering
\caption{Sensitivity to coefficient decomposition depth on SCALE}
\label{tab:depth}
\resizebox{\columnwidth}{!}{%
\begin{tabular}{|l|c|c|c|c|c|c|}
\hline
\multirow{2}{*}{\thead{Depth}} & \multicolumn{3}{c|}{\thead{Error-bound mode}} & \multicolumn{3}{c|}{\thead{PSNR mode}} \\
\cline{2-7}
 & \thead{\#best} & \thead{Avg.} & \thead{Worst} & \thead{\#best} & \thead{Avg.} & \thead{Worst} \\
\hline
Two-level   & 0  & 1.63\% & 5.00\% & 0  & 1.13\% & 4.16\% \\
\hline
Three-level & \textbf{{11}} & \textbf{{0.14\%}} & \textbf{{0.81\%}} & \textbf{{13}} & \textbf{{0.13\%}} & \textbf{{0.83\%}} \\
\hline
Four-level  & 5  & 0.55\% & 3.63\% & 3  & 2.06\% & 3.71\% \\
\hline
\end{tabular}
}
}
\end{table}

\subsection{Performance optimization}\label{sec:opt}
While our major goal is to improve the quality of progressive compression, keeping reasonable throughput is equally important, especially for data transfer tasks when refactoring/reconstruction operations lie on the critical paths.  
This section details our efforts to optimize the performance of the proposed framework. 

\paragraph{Fastest direction interpolation}
Inspired by recent work~\cite{liu2024high}, we found that all interpolations can be enforced to be performed along the fastest-varying direction for better cache efficiency and thus higher performance. 
In particular, we observe that the per-level linear interpolation described in \cref{sec:interpolation} is mathematically equivalent to sequentially interpolating the three regions within a level: first predicting round-type points along the $x$-direction using 2 surrounding triangle vertices, then predicting star-type points along the $y$-direction using 2 already-updated round points, and finally predicting diamond-type points along the $z$-direction using 2 already-updated star points (see \cref{fig:interp} for reference).
This reformulation does not change the interpolation directions or the predicted values; it only changes the order in which data points are visited.
Crucially, each region can be traversed along the fastest-varying dimension (i.e., the contiguous dimension in memory), ensuring that all interpolation steps remain cache-friendly without any data reordering.
This strategy also applies to per-region interpolation: since each region is interpolated independently, each region naturally corresponds to its own buffer, and the traversal within each region follows the fastest direction as well.

\paragraph{Optimized generic bitplane encoding}
As noted in \texttt{PMGARD}~\cite{liang2021error}, generic bitplane encoding is relatively slow despite its better efficiency in the error-bound mode, because the encoding operation has a branch to check if any bit has been stored for the current data. 
While this is necessary to ensure correctness and efficiency, it is no longer needed when the sign bits for all data points are recorded. 
As such, we decompose the algorithm into two sections: the first section retains the original branch when at least one data point has not recorded its sign, and it switches to the second section when all signs are recorded. 
In the second section, we eliminate such a branch to accelerate the encoding process.
This yields a substantial speedup over the original per-bit encoding implementation, since most data chunks complete the sign recordings with only a few bitplanes.  


\begin{figure}[htbp]
    \centering
    \includegraphics[width=\linewidth]{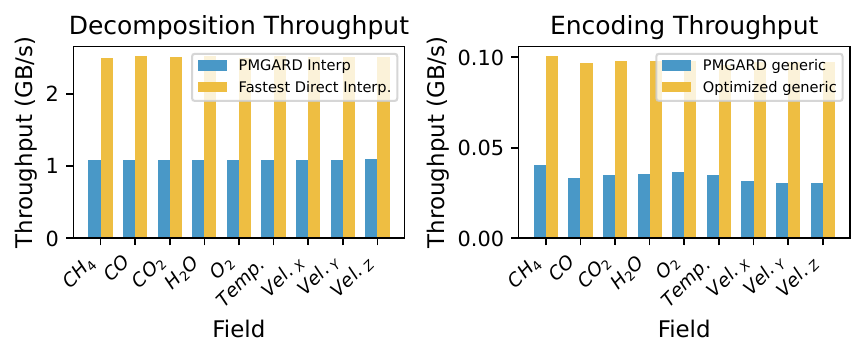}
    \caption{Throughput of decomposition and generic bitplane encoding before and after optimization on the S3D dataset with 9 valid fields}
    \label{fig:opt_throughput}
\end{figure}

We validate the proposed optimization using all 9 fields in the S3D dataset and present the results in \cref{fig:opt_throughput}. It is observed that both optimizations yield a $2\times$ performance improvement over existing implementations for the corresponding operations. 

\section{Evaluation}
\label{sec:evaluation}
We evaluate our method, named \texttt{ProAICD} for \underline{P}rogressive compression with \underline{A}daptive \underline{I}nterpolation and \underline{C}oefficient \underline{D}ecomposition, and compare it with three state-of-the-art approaches: \texttt{PMGARD}~\cite{liang2021error}, \texttt{SZ3-R}~\cite{magri2023general}, and \texttt{IPComp}~\cite{psz2025hpdc} in terms of retrieval efficiency, reconstruction quality, and throughput using five real-world datasets.
We also present two typical scientific use cases for end-to-end data transfer and snapshot visualization.

\subsection{Experimental Setup}

\subsubsection{Benchmark datasets}
We evaluate on five real-world scientific datasets spanning multiple scientific domains: Climate Simulation (CESM~\cite{kay2015community}), Hydrodynamics Simulation (Miranda~\cite{miranda}), Weather Simulation (SCALE~\cite{SCALE}), Combustion Simulation (S3D~\cite{chen2017s3d}), and large-scale Turbulence Simulation (JHTDB~\cite{rotstrate4096}).
The information about these datasets is detailed in Table~\ref{tab:data}.
Note that we exclude the \textit{Temperature} field from SCALE and the \textit{N\textsubscript{2}} and \textit{Pressure} fields from S3D for the aggregated results, as \texttt{IPComp} encounters numerical overflow on these fields, producing a constant tiny bitrate, no error guarantee, and negative PSNR, 
and we further report the results on excluded fields separately without \texttt{IPComp}.
We only evaluate JHTDB in the parallel data transfer experiments due to its large size.

\begin{table}[htbp]
{
\footnotesize
\centering
\caption{Datasets}
\label{tab:data}
\footnotesize
\resizebox{\columnwidth}{!}{%
\begin{tabular}{|l|c|c|c|c|}
\hline
\thead{Dataset} & \thead{Dimensions} & \thead{Valid Fields} & \thead{Type} & \thead{Size}\\ 
\hline
CESM & $26\times 1800\times 3600$ & 33 & double & 41.42 GB\\
\hline
Miranda & $256\times 384\times 384$ & 7 & double & 1.97 GB\\
\hline
SCALE & $98\times 1200\times 1200$ & 11 & double & 11.57 GB\\
\hline
S3D & $500\times 500\times 500$ & 9 & double & 8.38 GB \\
\hline
JHTDB & $4096\times4096\times4096$ & 1 & double & 512 GB \\
\hline
\end{tabular}
}
}
\end{table}

\subsubsection{Platform}
All experiments are conducted on the Morgan Compute Cluster (MCC)~\cite{mcc}, a medium-scale cluster with 100 Gbps InfiniBand HDR interconnect. 
Each compute node in the system is equipped with 2 AMD EPYC ROME 7702P processors, each with 64 cores and 256 GB of memory. Experiments associated with runtime are evaluated three times, and the average number is reported. 

\subsubsection{Quality assessment}
We assess the efficiency of progressive retrieval using the widely adopted rate-distortion graph~\cite{tao2019optimizing, jiao2022toward, wu2024error}, which depicts the relationship between bit-rate and distortion.
Bit-rate represents the average number of bits per data point in the retrieved data and serves as the x-axis, computed as $S_{\tau} \times 8 / n$, where $S_{\tau}$ is the retrieved size under requested tolerance $\tau$ and $n$ is the total number of data points. 
We include the metadata size in $S_{\tau}$ for all compressors, since metadata must be loaded to interpret retrieval sizes. 
To evaluate our two optimization targets, we use two types of rate-distortion plots: bit-rate versus relative error bound, and bit-rate versus PSNR.
In the former, curves that are lower and farther left indicate higher efficiency; in the latter, curves that are higher and farther left indicate higher efficiency.

Since \texttt{PMGARD} utilizes different encoding methods for the two targets, we use the generic bitplane encoding for \texttt{PMGARD} in the bit-rate versus error bound graphs and negabinary encoding in the bit-rate versus PSNR graphs.
\texttt{SZ3-R} iteratively compresses the residuals into 18 snapshots with target error bounds from $10^{-1}$ to $10^{-18}$ to fully preserve the precision limits of double-precision data.

Note that bit-rate and PSNR in \cref{fig:baseline_eb} and \cref{fig:baseline_psnr} are aggregated using all fields from the same dataset due to the limited space. 
The aggregated bit-rate is easily computed as the average bit-rate of all fields since they all share the same size in one dataset. The aggregated PSNR is computed as $20\log_{10}\frac{\sqrt{n_f}}{\sum_{j=1}^{n_f}NRMSE_{j}^2}$, where $n_f$ stands for number of fields in one dataset, and $NRMSE_j=\frac{RMSE_j}{\max(\bm{X_j})-\min(\bm(X_j))}$ for $j$-th field.

\subsection{Ablation Study}
We take the S3D dataset with 9 fields for the ablation study to demonstrate our efficiency gain step by step.

For error-bound mode, illustrated in \cref{fig:ablation_eb}, we use \texttt{PMGARD} as the baseline because per-level interpolation originates from \texttt{PMGARD}. Adaptive interpolation (AdatInterp) consistently improves efficiency over \texttt{PMGARD} across all nine fields.
Coefficient decomposition (CoeffDecom) further widens this gap, especially for \textit{CH\textsubscript{4}}, \textit{CO}, \textit{CO\textsubscript{2}}, \textit{H\textsubscript{2}O}, \textit{O\textsubscript{2}}, and \textit{Temperature}.
At worst, it preserves efficiency for \textit{Velocity\textsubscript{X}},
\textit{Velocity\textsubscript{Y}}, and \textit{Velocity\textsubscript{Z}} rather than degrading it.

As for PSNR mode, as shown in \cref{fig:ablation_psnr}, we take \texttt{IPComp} as the baseline since \texttt{IPComp} first applies per-region interpolation into the progressive method. 
According to \cref{fig:ablation_psnr}, we observe that AdatInterp still improves the efficiency by exploring the best-fit interpolation order while \texttt{IPComp} fixes the order, and then CoeffDecom obviously further improves the data quality throughout all nine fields.



\begin{figure}[htbp]
    \centering
    \includegraphics[width=\linewidth]{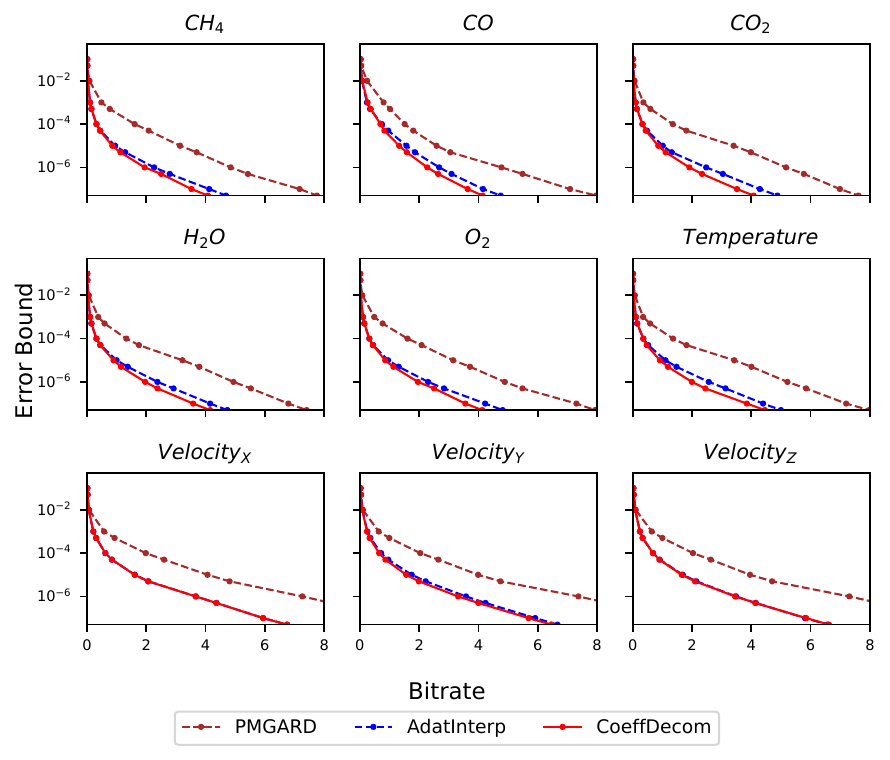}
    \vspace{-2.5em}
    \caption{Ablation study of error-bound mode on the S3D dataset.}
    \label{fig:ablation_eb}
\end{figure}

\begin{figure}[htbp]
    \centering
    \includegraphics[width=\linewidth]{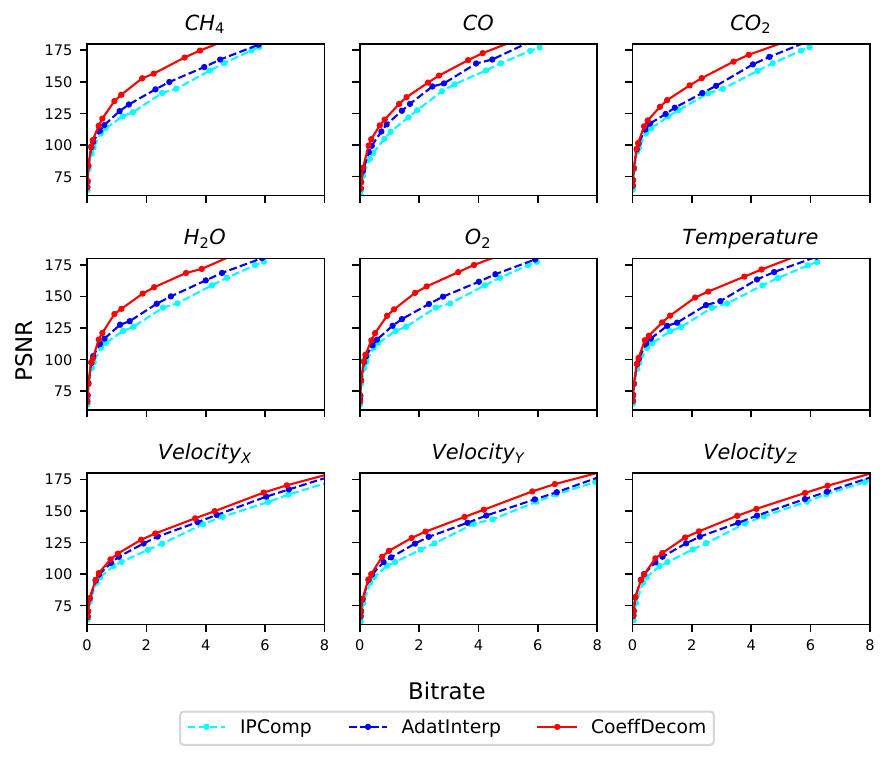}
    \vspace{-2.5em}
    \caption{Ablation study of PSNR mode on the S3D dataset.}
    \label{fig:ablation_psnr}
\end{figure}

\subsection{Comparison with State of the Arts}
\label{sota}
We then compare our method with 3 state-of-the-art progressive methods as mentioned earlier.

\subsubsection{Efficiency}
We present our improvement in error-bound mode in \cref{fig:baseline_eb} and in PSNR mode in \cref{fig:baseline_psnr} using all valid fields of the first four datasets in \cref{tab:data}.
Results on the three aforementioned excluded fields are presented in \cref{fig:baseline_eb_excluded} and \cref{fig:baseline_psnr_excluded}.

For error-bound mode, as shown in \cref{fig:baseline_eb}, compression ratios are improved by up to $23.9\%$, $29.7\%$, $29.3\%$, and $42.3\%$ respectively, when compared with the best-performing methods between \texttt{PMGARD}, \texttt{SZ3-R}, and \texttt{IPComp}. 
\texttt{ProAICD} outperforms \texttt{PMGARD} and \texttt{IPComp} across all error bounds on all datasets.
\texttt{SZ3-R} achieves a lower bit-rate at certain relatively loose error bounds on the CESM and SCALE datasets; however, this is because those error bounds coincide with the exact target error bounds used in its residual-based compression, giving it a natural advantage at those specific operating points.
Even though tighter error bounds such as $10^{-6}$ and $10^{-7}$ are also directly targeted by \texttt{SZ3-R}, this advantage diminishes as the error bound decreases, due to the inherent redundancy in its residual-based compression scheme.
We can also observe similar advantages on excluded fields from ~\cref{fig:baseline_eb_excluded}.

For PSNR mode, as illustrated in \cref{fig:baseline_psnr}, compression ratios are further optimized by up to approximately $92.5\%$, $51.1\%$, $75.6\%$, and $91.3\%$, respectively, compared with the best-performing methods.
Note that unlike non-progressive methods such as~\cite{jiao2025qp}, where a given error bound deterministically produces a specific PSNR, progressive methods retrieve coefficients incrementally, so the same requested error bound does not guarantee the same PSNR across different methods.
To enable fair comparison, we apply linear regression between points on each rate-distortion curve and compare bit-rates at matched PSNR values.
Our method clearly outperforms all baselines across all datasets (including the excluded fields shown in ~\cref{fig:baseline_psnr_excluded}) in terms of efficiency in PSNR mode.

\begin{figure}[htbp]
    \centering
    \includegraphics[width=\linewidth]{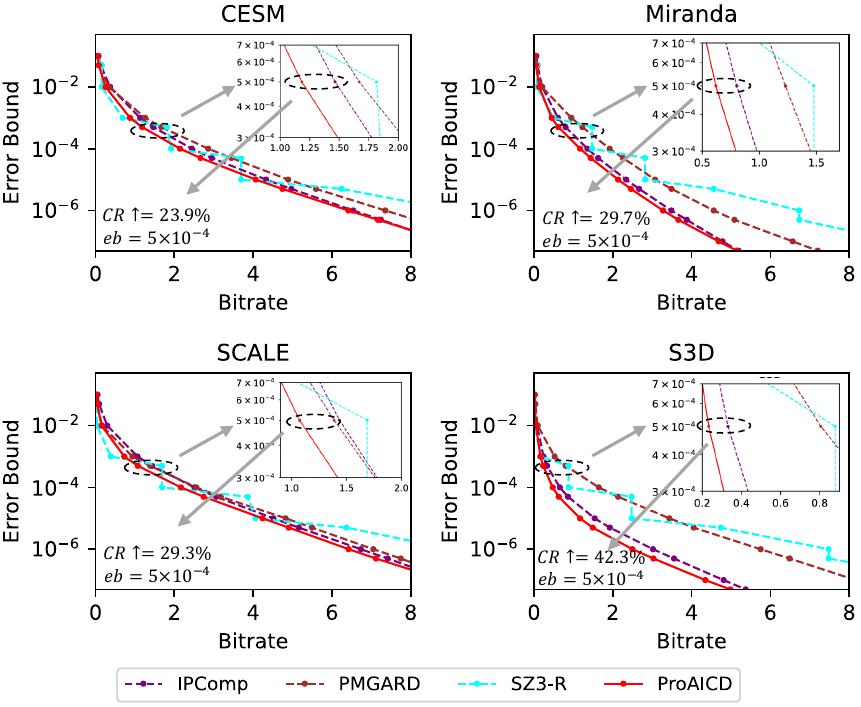}
    \caption{Baseline comparison of error-bound mode.}
    \label{fig:baseline_eb}
\end{figure}

\begin{figure}[htbp]
    \centering
    \includegraphics[width=\linewidth]{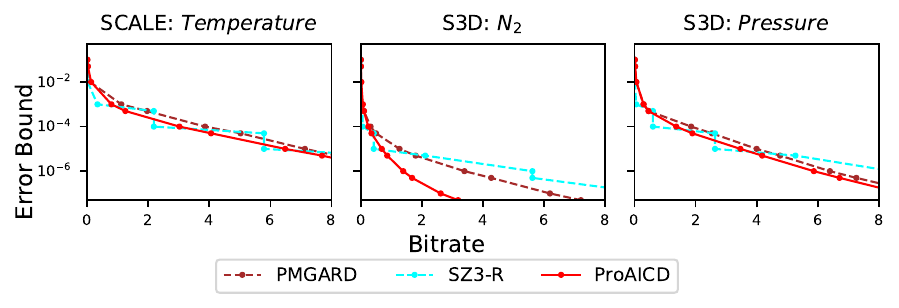}
    \caption{Baseline comparison of error-bound mode on excluded fields.}
    \label{fig:baseline_eb_excluded}
\end{figure}

\begin{figure}[htbp]
    \centering
    \includegraphics[width=\linewidth]{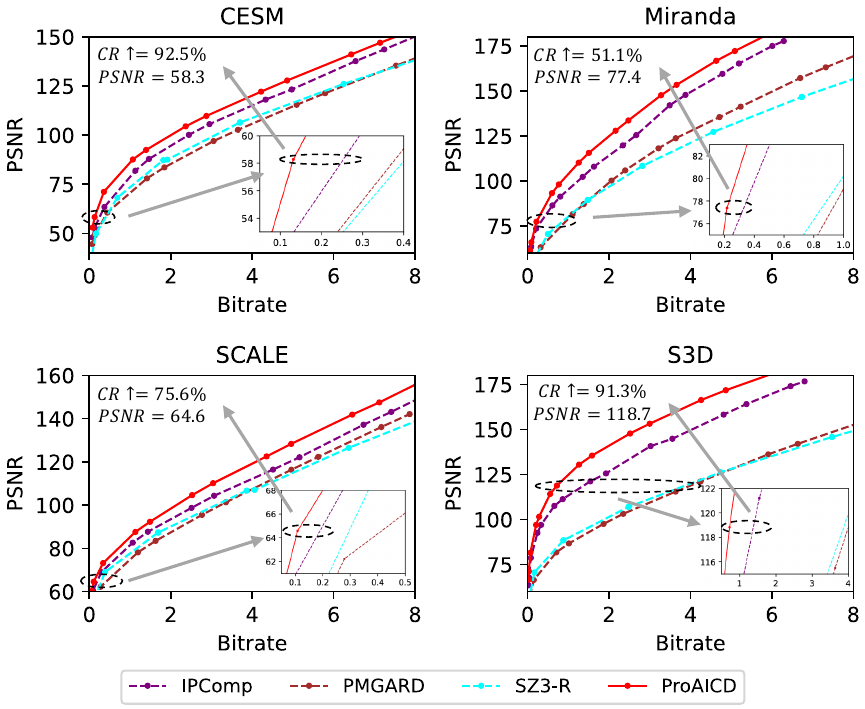}
    \caption{Baseline comparison of PSNR mode.}
    \label{fig:baseline_psnr}
\end{figure}

\begin{figure}[htbp]
    \centering
    \includegraphics[width=\linewidth]{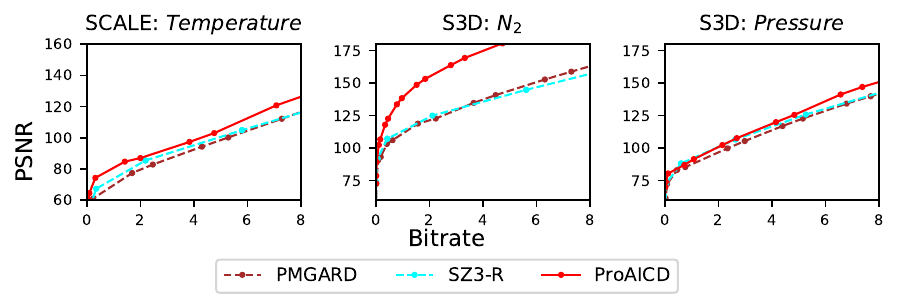}
    \caption{Baseline comparison of PSNR mode on excluded fields.}
    \label{fig:baseline_psnr_excluded}
\end{figure}

\subsubsection{Performance}

We report the average refactor time and average reconstruction time of the four methods in \cref{tab:refactor_time} and \cref{tab:reconstruct_time}, using three requested relative error bounds of $10^{-2}$, $10^{-4}$, and $10^{-6}$ as an example for reconstruction.

For refactoring, \texttt{PMGARD}'s PSNR mode is the fastest across all four datasets, followed by \texttt{IPComp}.
Our PSNR mode achieves comparable refactoring time to \texttt{IPComp}, while our error-bound mode is moderately slower due to the additional overhead of tuning, coefficient decomposition, and ordering.
\texttt{SZ3-R} is the slowest by a significant margin, taking $4\times$--$10\times$ longer than the other methods, as it iteratively compresses the residuals at each target error bound from $10^{-1}$ down to $10^{-18}$.

For reconstruction, \texttt{IPComp} is the fastest at $\tau=10^{-4}$ and $10^{-6}$ on CESM, SCALE, and S3D.
Note that the reported time corresponds to a single retrieval request at the given tolerance rather than an accumulation over progressively refined requests.
Across all methods, reconstruction time increases as the tolerance tightens, since more data must be retrieved and decompressed; the increase is sharpest for \texttt{SZ3-R}, whose representation is a chain of residual snapshots, so a request at $\tau$ must retrieve and decompress every snapshot down to $\tau$, accumulating cost with the length of the chain rather than with the requested precision alone.
This trend holds across all three tolerances and all four datasets, confirming that it is not specific to a single operating point.
Separately, both \texttt{IPComp} and \texttt{SZ3-R} exhibit notably slow reconstruction on the Miranda dataset, despite it being the smallest dataset, and this ranking also holds across all three tolerances.
We observe a similar pattern in both single-machine and parallel experiments on the JHTDB dataset, suggesting that these two methods may not perform well on turbulence simulation datasets in general.


In terms of bit-rate, the comparison against \texttt{SZ3-R} is governed by the same accumulation. 
At the loose end, $\tau=10^{-2}$, only the first two components are needed, and it attains the lowest bit-rate on all four datasets, as a short residual chain is hard to beat when only a coarse approximation is requested. 
At $\tau=10^{-4}$, its advantage is already confined to CESM and SCALE, while our error-bound mode is the lowest on Miranda and S3D. 
Once the chain lengthens further, the accumulated cost of storing successive residuals outweighs the advantage entirely: at $\tau=10^{-6}$ our approach attains the lowest bit-rate on all four datasets, whereas \texttt{SZ3-R} becomes the most expensive, e.g., $BR=9.12$ versus $BR=6.42$ on CESM compared with \texttt{ProAICD}.

Overall, our error-bound mode achieves the lowest bit-rate on the majority of datasets at the cost of slightly longer refactoring and reconstruction time, while our PSNR mode maintains competitive speed with consistently lower bit-rate than all three baselines.
\begin{table}[htbp]
    \centering
    \caption{Average refactor time (in seconds) of different progressive approaches using all fields from the same dataset}
    \setlength{\tabcolsep}{3pt}
    \footnotesize
    \begin{tabular}{|c|c|c|c|c|c|}
        \hline
        \multicolumn{2}{|c|}{\multirow{2}{*}{\textbf{Method}}} & \multicolumn{4}{c|}{\textbf{Datasets}} \\
        \cline{3-6}
        \multicolumn{2}{|c|}{} & CESM & Miranda & SCALE & S3D \\
        \hline
        \multirow{2}{*}{\texttt{ProAICD}}   & EB   & 19.20 &  7.14 & 21.86 & 18.37 \\
        \cline{2-6}
                                          & PSNR & 17.05 &  3.96 & 14.63 & 14.49 \\
        \hline
        \multirow{2}{*}{\texttt{PMGARD}} & EB   & 32.41 &  9.90 & 27.24 & 33.54 \\
        \cline{2-6}
                                          & PSNR & \textbf{12.36} &  \textbf{2.82} & \textbf{10.36} &  \textbf{9.91}\\
        \hline
        \texttt{SZ3-R}                    & --   &106.59 & 42.51 & 93.18 & 81.48 \\
        \hline
        \texttt{IPComp}                   & --  & 14.77 &  4.77 & 12.17 & 11.39 \\
        \hline
    \end{tabular}
    \label{tab:refactor_time}
\end{table}

\begin{table}[htbp]
    \centering
    \caption{Average reconstruction time (in seconds) and average bit-rate (BR) of different progressive approaches using all fields from the same dataset under tolerances $\tau = 10^{-2}$, $10^{-4}$, and $10^{-6}$}
    \setlength{\tabcolsep}{3pt}
    \footnotesize
    \begin{tabular}{|c|c|c|c|c|c|c|c|c|c|}
        \hline
        \multicolumn{2}{|c|}{\multirow{2}{*}{\textbf{Method}}} & \multicolumn{2}{c|}{\textbf{CESM}} & \multicolumn{2}{c|}{\textbf{Miranda}} & \multicolumn{2}{c|}{\textbf{SCALE}} & \multicolumn{2}{c|}{\textbf{S3D}} \\
        \cline{3-10}
        \multicolumn{2}{|c|}{} & Time & BR & Time & BR & Time & BR & Time & BR \\
        \hline
        \multicolumn{10}{|c|}{$\tau = 10^{-2}$} \\
        \hline
        \multirow{2}{*}{\texttt{ProAICD}} & EB   &  4.09 & 0.25 &  0.98 & 0.14 &  3.32 & 0.15 &  3.43 & 0.05 \\
        \cline{2-10}
                                          & PSNR &  2.82 & 0.36 &  0.75 & 0.21 &  2.35 & 0.32 &  2.59 & 0.07 \\
        \hline
        \multirow{2}{*}{\texttt{PMGARD}}  & EB   &  4.04 & 0.34 &  0.78 & 0.20 &  2.96 & 0.16 &  1.80 & 0.10 \\
        \cline{2-10}
                                          & PSNR &  3.17 & 0.44 &  \textbf{0.68} & 0.32 &  2.41 & 0.28 &  1.95 & 0.15 \\
        \hline
        \texttt{SZ3-R}                    & --   &  \textbf{2.11} & \textbf{0.16} &  2.48 & \textbf{0.07} &  \textbf{1.71} & \textbf{0.04} &  \textbf{1.44} & \textbf{0.02} \\
        \hline
        \texttt{IPComp}                   & --   &  3.13 & 0.37 &  5.06 & 0.21 &  2.92 & 0.30 &  2.34 & 0.08 \\
        \hline
        \multicolumn{10}{|c|}{$\tau = 10^{-4}$} \\
        \hline
        \multirow{2}{*}{\texttt{ProAICD}} & EB   &  7.12 & 2.14 &  1.51 & \textbf{1.17} &  5.98 & 2.17 &  4.82 & \textbf{0.48} \\
        \cline{2-10}
                                          & PSNR &  3.55 & 2.36 &  0.92 & 1.26 &  2.99 & 2.48 &  3.06 & 0.56 \\
        \hline
        \multirow{2}{*}{\texttt{PMGARD}}  & EB   &  8.04 & 2.86 &  1.53 & 1.92 &  6.15 & 2.56 &  4.25 & 1.65 \\
        \cline{2-10}
                                          & PSNR &  4.33 & 3.05 &  \textbf{0.91} & 2.06 &  3.42 & 2.77 &  2.83 & 1.87 \\
        \hline
        \texttt{SZ3-R}                    & --   &  4.95 & \textbf{1.92} &  4.74 & 1.48 &  4.12 & \textbf{1.69} &  3.06 & 0.88 \\
        \hline
        \texttt{IPComp}                   & --   &  \textbf{3.14} & 2.45 &  5.06 & 1.36 &  \textbf{2.89} & 2.51 &  \textbf{2.27} & 0.66 \\
        \hline
        \multicolumn{10}{|c|}{$\tau = 10^{-6}$} \\
        \hline
        \multirow{2}{*}{\texttt{ProAICD}} & EB   & 12.21 & 6.42 &  2.28 & \textbf{3.28} & 10.25 & \textbf{6.43} &  7.22 & \textbf{2.50} \\
        \cline{2-10}
                                          & PSNR &  4.39 & \textbf{6.42} &  \textbf{1.10} & 3.28 &  3.65 & 6.43 &  3.69 & 2.54 \\
        \hline
        \multirow{2}{*}{\texttt{PMGARD}}  & EB   & 13.73 & 7.35 &  2.44 & 4.56 & 10.92 & 7.05 &  8.39 & 5.75 \\
        \cline{2-10}
                                          & PSNR &  5.37 & 7.53 &  1.12 & 4.71 &  4.30 & 7.17 &  3.58 & 5.91 \\
        \hline
        \texttt{SZ3-R}                    & --   & 11.29 & 9.12 &  8.46 & 6.73 &  9.02 & 9.02 &  7.36 & 7.48 \\
        \hline
        \texttt{IPComp}                   & --   &  \textbf{3.69} & 6.54 &  5.43 & 3.49 &  \textbf{3.29} & 6.73 &  \textbf{2.74} & 3.02 \\
        \hline
    \end{tabular}
    \label{tab:reconstruct_time}
\end{table}

\subsection{Use case for eb mode: end-to-end data transfer}
We demonstrate how our method can benefit scientific data management using a practical use case of remote data transfer. 
Here is the experiment setup: suppose the refactored data is stored at the Frontier Supercomputer at Oak Ridge Leadership Computing Facilities~\cite{frontier}, and the request for data is initiated from MCC~\cite{mcc} with a frequently used tolerance $\tau=10^{-4}$.
The data is transferred via Globus~\cite{foster1997globus}, the leading research cyberinfrastructure widely used for scientific data sharing and management.
This setup reflects a common scenario in which domain scientists must download data from remote servers to local machines for post hoc data analysis.
To illustrate how the progressive methods perform with increasing data sizes, we conduct a weak-scaling experiment using $128$, $256$, $512$, and $1024$ cores, respectively.
In all the experiments, each core is processing a fixed size ($256\times512\times512$) of data from the JHTDB dataset, so the total size of the processed data increases with the number of cores.
We plot the end-to-end data transfer time (which includes both retrieval time and transfer time) in \cref{fig:parallel}.

During retrieval, \texttt{SZ3-R} yields the smallest retrieved size because $\tau=10^{-4}$ is one of its targeted error bounds.
However, this advantage is offset by the longest retrieval time among all methods
as previously observed on the Miranda dataset, both \texttt{SZ3-R} and \texttt{IPComp} exhibit slow reconstruction on turbulence simulation datasets, and JHTDB falls into the same category.
Moreover, \texttt{IPComp} yields the second largest retrieved size, only smaller than \texttt{PMGARD}, further limiting its transferring performance.

Our method yields the second smallest retrieved size, only slightly larger than \texttt{SZ3-R}, mainly because it leverages the most correlation within the dataset using adaptive interpolation and coefficient decomposition techniques.
Meanwhile, our method achieves the second shortest retrieval time, only slightly slower than \texttt{PMGARD}.
As a result, our method takes the shortest end-to-end data transfer time in experiments with $256$, $512$, and $1024$ cores.
Specifically, performance gain is up to $1.26\times$ over the best existing method and $18.17\times$ over vanilla transfer of the original data.

\begin{figure}[t]
    \centering
    \includegraphics[width=\linewidth]{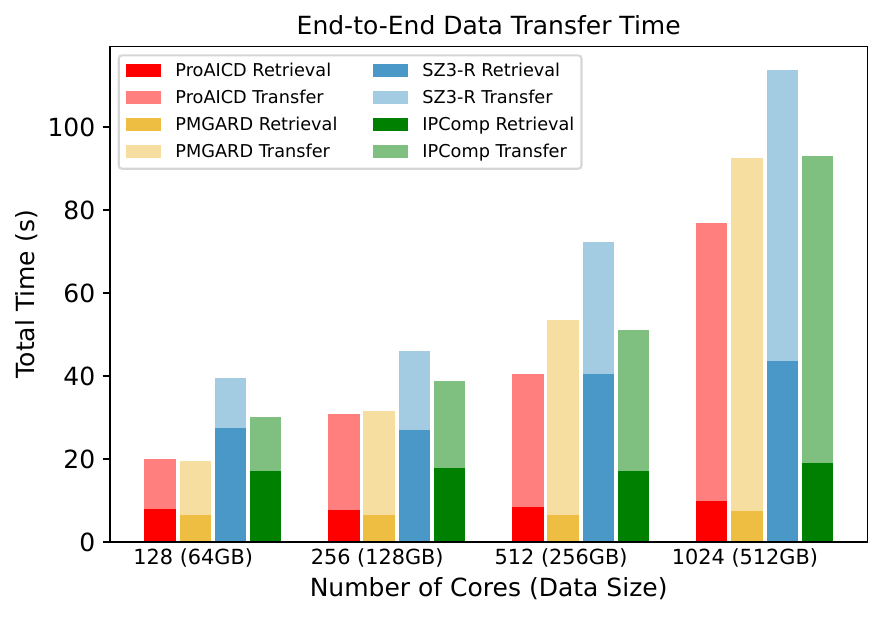}
    \caption{End-to-end data transfer time using JHTDB dataset. Transferring the original data of 128, 256, 512, and 1024 cores takes 189, 373, 735, and 1342 seconds, respectively.}
    \label{fig:parallel}
\end{figure}

\subsection{Use case for PSNR mode: data visualization}

We illustrate the practical impact of progressive retrieval on scientific data visualization in \cref{fig:visual}, using the \textit{Temperature} field from the CESM dataset as an example.
For the remaining progressive methods, we cap the retrieval bitrate at approximately 0.5 to ensure a fair comparison.
Notably, our method produces the most faithful reconstruction at a bitrate of only 0.37 and is visually nearly indistinguishable from the original, whereas the other methods show noticeable artifacts or over-smoothing even at higher bitrates.
In other words, our method retrieves substantially less data from disk yet yields better visual quality, a property that is especially valuable in interactive exploration of large-scale scientific datasets.


\begin{figure}[h]
    \centering
    \includegraphics[width=\linewidth]{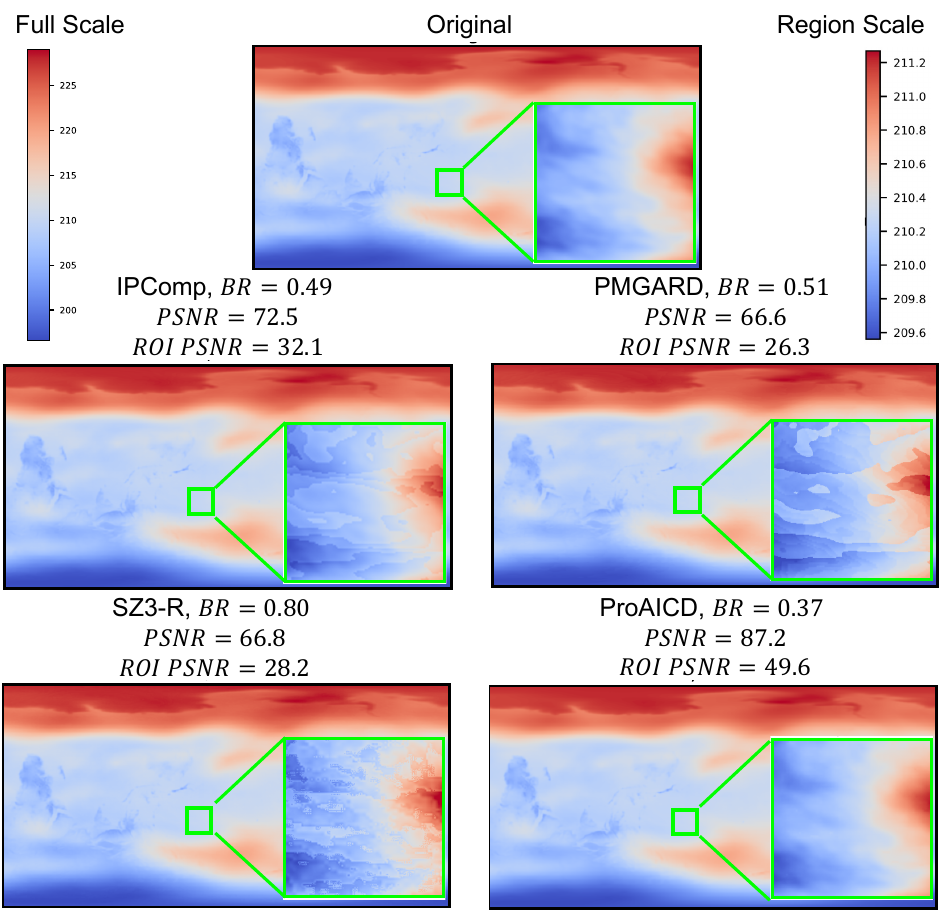}
    \caption{Visual comparison of reconstructed data from different progressive approaches on the CESM dataset, field \textit{Temperature}, at the slice [11, 1000:1200, 1480:1680]. Each panel shows the method name, achieved bit-rate, global PSNR over the full field, and ROI PSNR computed within the displayed region. The left and right colorbars represent the color scales in the full slice and the displayed region, respectively.}
    \label{fig:visual}
\end{figure}

\section{Conclusion}
\label{sec:conclusion}

In this paper, we design an adaptive progressive compression framework to enable efficient retrieval of scientific data toward diverse targets.
Our approach improves both retrieval efficiency and reconstruction quality by introducing a highly adaptive design with respect to interpolation schemes and encoding methods, and by employing a novel coefficient decomposition method that exploits the commonly overlooked correlation among decorrelated coefficients.

Experimental evaluations demonstrate that the error-bound mode of our method delivers up to $42.3\%$ improvement in compression ratio with comparable throughput over state-of-the-art approaches under the same requested tolerance, leading to up to $1.26\times$ speedup in end-to-end data transfer.
The PSNR mode further achieves up to $92.5\%$ improvement under the same PSNR with comparable throughput, and produces the best visualization quality with the least retrieved data.

In future work, we plan to explore novel algorithms to further exploit correlations within both raw data and decorrelated coefficients, and to extend our framework to GPUs to further improve throughput.

\section*{Acknowledgment}
The research is supported in part by the U.S. Department of Energy (DOE) RAPIDS-3 SciDAC and Sirius-2 projects under contract number DE-AC05-00OR22725, and National Science Foundation (NSF) under Grant OAC-2628470, OAC-2628471,  OAC-2628472, OAC-2311757, and OAC-2144403. This research used resources of the Oak Ridge Leadership Computing Facility (OLCF), which is a DOE Office of Science User Facility. 

\bibliographystyle{IEEEtran}
\bibliography{references}

\end{document}